# Experimental determination of the glass transition temperature in a very narrow temperature range by Temperature Modulated Optical Refractometry

A. Klingler, B. Wetzel, J. K. Krüger*

Leibniz-Institut für Verbundwerkstoffe GmbH (IVW), RPTU Kaiserslautern-Landau, Erwin-Schrödinger-Straße 58, 67663 Kaiserslautern, Germany

Corresponding author: A. Klingler: andreas.klingler@ivw.uni-kl.de, J. K. Krüger: jan-kristian.krueger@ivw.uni-kl.de

**Abstract**

Latest since the landmark studies of Kovacs and co-workers on the glass transition of polymers, it is clear that thermally induced volume changes are of central importance for the understanding of the nature of the glass transition. Due to the kinetic background of the canonical (thermal) glass transition, it does not seem possible to derive a well-defined glass transition temperature $T_g$ based on susceptibilities such as the thermal volume expansion coefficient, $\beta(T)$, being strongly coupled to the glass transition process. Therefore, in practice, $T_g$ is for example defined via the inflection point of the step-like $\beta(T)$ curve. In this publication, we propose to use a thermo-optical feature, preceding the glass transition in the high-temperature phase, to determine the glass transition temperature $T_g$ of a model polymer in a rather narrow temperature interval.

**Main text:**

*As a preliminary note*: the present publication is not concerned with a further clarification of the nature of the glass transition, but with the question if the so-called thermo-optical coefficient (TOC) [1] is a suitable sensor for determining the temperature at which a polymeric glass former reaches solid-state properties in the cooling mode. This temperature could then be called the glass transition temperature $T_{g,TOC}$. The following investigations are carried out on a model epoxy, abbreviated as *EP*, being based on a mixture of diglycidyl ether of bisphenol A (DER332, Merck KGaA

(Germany), CAS 1675-54-3, molar mass 340.41 g/mol) and difunctional (23 wt.-%) and trifunctional (77 wt.-%) carboxylic acids (Pripol™ 1040, Croda (United Kingdom)). Samples were prepared in a stoichiometric ratio of epoxy to carboxylic acid moieties and cured by the addition of 10 mol-% of TBD (1,5,7-Triazabicyclo[4.4.0]dec-5-en, Merck KGaA (Germany), CAS 5807-14-7, relative to carboxylic acid functions). The cured EP has a glass transition temperature of about $T_{g,17mHz} = 13°C$.

The glass transition of canonical glass formers is isostructural in nature. Both, the high temperature and low temperature phases have isotropic symmetry. The most important susceptibilities coupling to the glass transition are, besides the thermal volume expansion $\beta$, the specific heat $c_p$ and the shear-elastic constant $G$ [2,3].

Due to the kinetic character of the canonical glass transition, the susceptibilities that couple to the glass transition, on slow cooling, show a smooth, kink-like step without any distinctive feature that would be suitable for defining a transition temperature $T_g$ [4]. Therefore, a unique definition of a glass transition temperature based on a susceptibility anomaly accompanying the glass transition seems impossible. As a consequence, the inflection point of a susceptibility anomaly reflecting the glass transition is often defined as an "operative" glass transition [2,5–7]. Alternatively, the temperatures at the onset or offset of slope changes of measured susceptibilities at the glass transformation process are chosen, respectively [7]. In the case of dynamically measured susceptibilities, the temperature at which the loss maximum occurs is defined as the dynamic glass transition temperature, $T_{g,dyn}$, for a given measurement frequency $f = \omega/2\Pi$ [5,6]. For the sake of completeness, two virtual glass transition temperatures should be mentioned: The Vogel-Fulcher-Tamman (VFT) temperature $T_{VFT}$ and the Kauzmann temperature $T_K$. $T_{VFT}$ is the temperature at which the so-called α-relaxation time would diverge due to the dynamics of the glass transition process, if the α-relaxation process would not already be stopped at 30°C-40°C above $T_{VFT}$ through the quasi-static glass transition [8–10]. With the virtual Kauzmann temperature $T_K$ the excess entropy would become negative [3,11,12]. Also, the temperature $T_K$ is generally about 30°C-40°C below the experimentally determined quasi-static $T_g$.

Thus, if one cannot derive a glass transition temperature from the susceptibility anomalies of a canonical glass transition itself, the question arises whether this is possible by freezing a property that is causally independent of the glass formation process. The measurement of additional dynamic thermo-optical properties using "Temperature Modulated Optical Refractometry" (TMOR) allows this.

TMOR enables static and dynamic measurements of the thermo-optical coefficient, $\Psi^*(\omega,T)$ with $\omega = 2\Pi f$ [1,13].

$$\Psi^* = \frac{\partial n_\omega^*}{\partial T} = \frac{A_n}{A_T} e^{i\Phi} \qquad \text{[Eq. 1]}$$

where $n_\omega^*$ is the complex refractive index response, $A_n$ is the amplitude of the refractive index response due to the temperature perturbation and $\Phi$ is the phase lag between the sinusoidal temperature perturbation and the refractive index response.

Different physical or chemical processes can affect the thermo-optical coefficient. The most prominent influencing variable is probably the static and dynamic thermal volume expansion, from which the static and dynamic coefficient of thermal volume expansion, $\beta_{stat}(T)$ and $\beta^*(t,T,f) = \beta'(t,T,f) + i\beta''(t,T,f)$ can be derived [14,15].

$$\beta'(t,T,f) = \frac{-6(N_{\text{mean}}(t,T))}{[N_{\text{mean}}^2(t,T)-1][N_{\text{mean}}^2(t,T)+2]} \frac{A_n(t,T)}{A_T} \cos(\Phi(t,T,f)), \qquad \text{[Eq. 2]}$$

$$\beta''(t,T,f) = \frac{-6(N_{\text{mean}}(t,T))}{[N_{\text{mean}}^2(t,T)-1][N_{\text{mean}}^2(t,T)+2]} \frac{A_n(t,T)}{A_T} \sin(\Phi(t,T,f)). \qquad \text{[Eq. 3]}$$

where $N_{mean}$ is the refractive index average over one modulation period, $A_T$ is the amplitude of the temperature modulation, $A_n$ is the refractive index response to $A_T$ and $\Phi$ is the phase angle between $n(t)$ and $T(t)$.

In the absence of dynamics, or for very low modulation frequencies, $\beta'$ yields in a good approximation the static thermal volume expansion coefficient $\beta_{stat}$. An exhaustive description of the TMOR theory and background can be found elsewhere [1].

This also applies to the viscoelastic state and the transition to the glassy state of the *EP*, used in this study. Since the coefficients of thermal volume expansion ($\beta_{stat}(T)$ and $\beta^*(f,T)$) are the physical quantities causally controlling the canonical glass transition, these quantities can have an important but only secondary significance for our more precise definition of a quasi-static glass temperature $T_g$. Here, the authors are interested in a contribution to the thermo-optical coefficient due to an unexpected interaction of the *EP* with the optical prism of the TMOR device, due to the temperature modulation. This aspect has partially been addressed in [13], already. When the *EP* is placed in the TMOR device at elevated temperatures ($T \gg T_{g,dyn} \sim 10°C$), the *EP* was found to strongly adhere to the measuring prism surface. This means, the first molecular layer of *EP* is fixed at the prism surface. Here, it is important to point out that the thermal expansion coefficient of the prism material (e.g. YAG, $\beta_{YAG,300K} = 2.4 \cdot 10^{-5}/K$) is about 20 times smaller than the thermal expansion coefficient of our *EP* sample [16]. In other words, when the temperature is modulated in the TMOR device, the first molecular sample layer adhering to the prism remains unaffected by the continuous temperature changes. Consequently, the temperature modulation generated at the prism surface and the adjacent sample creates a dynamic inhomogeneous and anisotropic mechanical field in the sample. At larger distance from the prism surface, this field transforms to a dynamic deformation field of pure dynamic thermal volume expansion. At lower temperatures, when the glass transition is approached, molecular translational motions, generated by this field, become frozen and vanish. This stress field situation is schematically given in Figure *1*.

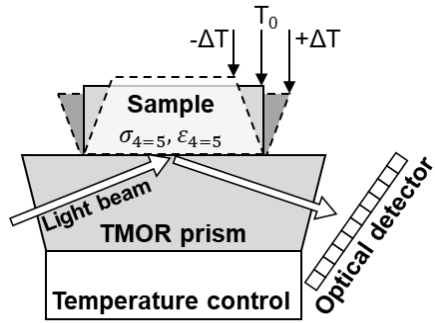

Figure 1: Schematic shear-field situation in the EP sample. The collected data originates approximately from a small cylindrical volume of a diameter of 1mm and a height of ~250nm.

If both features, i.e. the dynamic thermal volume expansion and the shear deformation, are active in the information-volume of TMOR, both will contribute to the thermo-optical property, and can thus be denoted separately. The complex contribution originating from the thermal expansion is given by

$$\beta^*_{thex}(f,T) = (\beta'_{thex}(f,T), \beta''_{thex}(f,T)), \quad \text{[Eq. 4]}$$

and the additive but independent contribution in the viscoelastic phase to $\beta^*(T,f)$ is given as

$$\beta^*_{add}(f,T) = (\beta'_{add}(f,T), \beta''_{add}(f,T)). \quad \text{[Eq. 5]}$$

Figure 2 shows the complex thermal volume expansion coefficients $\beta'_{thex}(f,T)$ and $\beta''_{thex}(f,T)$ calculated from the dynamic thermo-optical coefficient $\Psi^*(f,T)$ according to Eq. 2 and Eq. 3 for two modulation frequencies 17mHz and 1.7mHz. Thus, these calculations were performed under the assumption that the thermo-optical coefficient $\Psi^*(f,T)$ was affected only by thermal expansion.

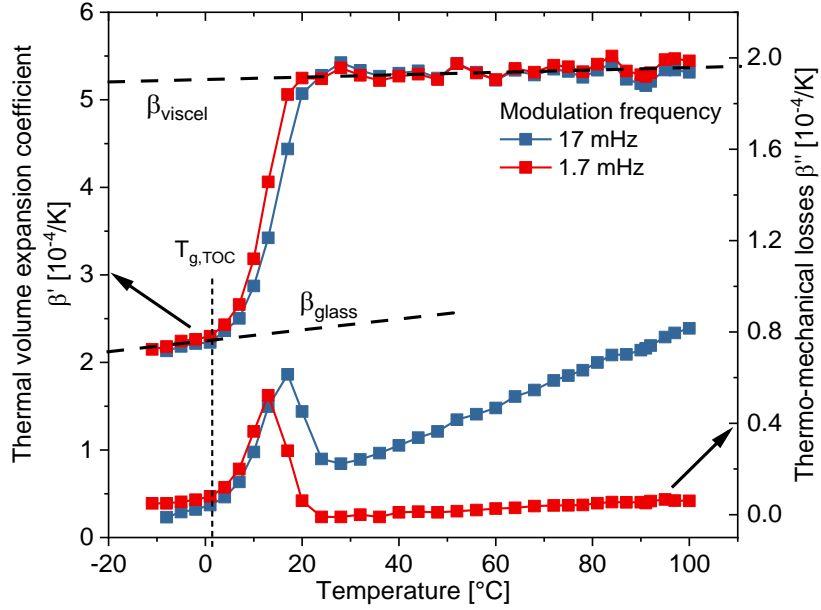

Figure 2: Thermal volume expansion and thermo-mechanical losses as a function of Temperature

For the measurement frequency $f = 1.7mHz$, the assumptions that have been made seem to be correct. $\beta'_{thex}(T)$ and $\beta''_{thex}(T)$ show the classical temperature behaviour associated with the so-called α-process: $\beta'_{thex}(T)$ reflects the dynamic glass transition by a step-like anomaly and $\beta''_{thex}(T)$ shows a loss maximum at the inflection point of $\beta'_{thex}(T)$. The inflection point and the loss maximum are slightly above the quasi-static $T_g$ for the given *EP* and are interpreted as the dynamic glass transition temperature $T_{g,dyn}$ for the given probe frequency of $f = 1.7mHz$. Within the margin of error, $\beta'_{thex}(T)$ behaves almost horizontal well above $T_{g,dyn}(f = 1.7mHz)$ and, as expected, $\beta''_{thex}(f,T)$ is practically zero outside the relaxation region.

According to Figure 2 the situation changes when the temperature modulation frequency is increased by a factor of 10 to $f = 17mHz$. Then, the entropy production reflected by $\beta''_{add,17mHz}(T)$ now increases linearly with the temperature in the viscoelastic phase of *EP*, while the real part $\beta'_{add,17mHz}(T)$ coincides with

$\beta'_{thex,17mHz}(T)$. This raises the question of the physical nature of the additional losses, which, moreover, increase with rising temperature and, at 100°C, clearly exceed those of the maximum loss due to dynamic thermal expansion at about 19°C. This drastic increase of $\beta''_{add,17mHz}(T)$ with simultaneous constancy of $\beta'_{add,17mHz}(T)$ clearly questions the Kramers-Kronig relationship [17–19].

The fact that the Kramers-Kronig relationship does not hold above $T_g$ indicates the extensive independence of the influence of thermal expansion and the observed additional influence on the thermo-optical coefficient in the viscoelastic region of the *EP*.

In the following, having extended the TMOR measurements presented in Figure 2 by additional temperature modulation frequencies, it will be shown that this detection method yields a well-defined glass transition temperature $T_g$ for the *EP* sample.

Figure 3 shows 10 low frequency TMOR measurements in a frequency range of $f = 1.7mHz$ to $f = 17mHz$. All $\beta''(T)$ curves show the loss peak caused by the α-relaxation process. As expected, the maximum of the loss peak shifts to lower temperatures with decreasing temperature modulation frequency. The losses $\beta''_{add}(f,T)$, additional to those induced by thermal volume expansion, in the viscoelastic temperature range, decrease linearly with decreasing temperature for each of the temperature modulation frequencies. The slopes of these curves systematically decrease with decreasing frequency.

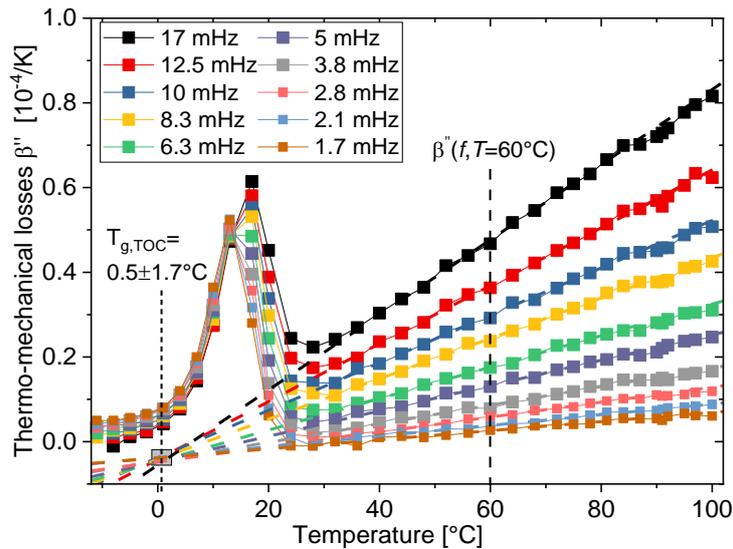

Figure 3: Low frequency measurements of EP showing the frequency-dependent shift of the glass transition (peak in $\beta''$) and the unexpected temperature-modulation induced losses at $T > T = 30°C$

Exemplarily extracting $\beta''_{add,60°C}(f)$ from Figure 3 and plotting it as a function of the temperature modulation frequency, we again find surprisingly a linear $\beta''$- behavior (Figure *4*). Figure *4* clearly shows that the temperature modulation is the cause of the additional losses, because $\beta''_{add,60°C}(f)$ completely disappears with the modulation frequency ($f \to 0 Hz$).

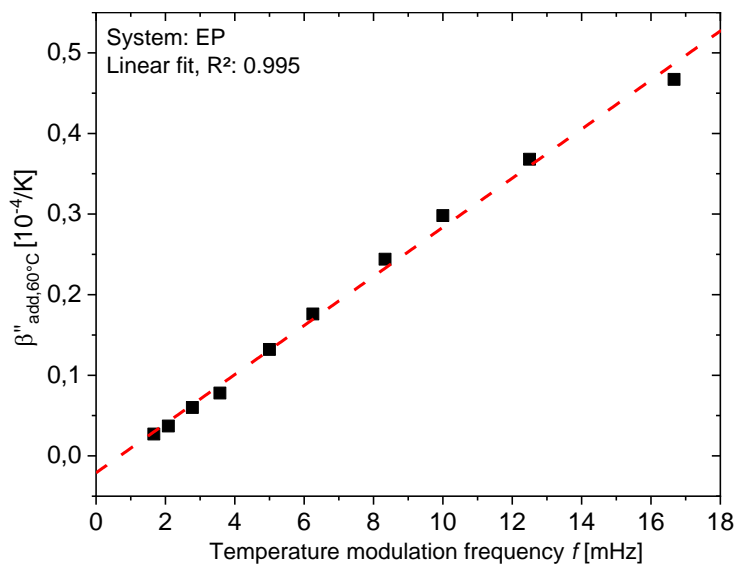

Figure 4: Additional thermo-optical losses as a function of the temperature modulation frequency in the high temperature phase at $T = 60°C$. Red, dashed line: least squares fit

Coming back to Figure 3, an alternative perspective is provided by the linear parts of the loss curves $\beta''_{add,f}(T)$ in the high temperature phase of our *EP on the* adjacent glass transition region. The given dashed line fits are based for each of the given temperature modulation frequencies $f$ on a least squares fit, as a function of temperature T. Only the measured data from the viscoelastic region ($T > 40°C$) were used for the fits. Thus, in the glass transition region, these straight lines are extrapolations and represent the remaining losses $\beta''_{add,f}(T)$, which interfere with the thermal strain losses $\beta''_{thex,f}(T)$ in the glass transition region. As the temperature decreases and the glass transition is approached, $\beta''_{add,f}(T)$ decreases linearly and runs into a minimum of $\beta''_{add,f}(T)$ (~0/K). In the vicinity of this $\beta''_{add,f}(T)$ -minimum, all straight line fits intersect as a function of the parameter "temperature modulation frequency" $f$. In other words, the dynamic, anisotropic and inhomogeneous deformation of *EP* triggered by the temperature modulation, vanishes with the solidification of the *EP* sample.

The glass transition temperature determined in this way is $T_{g,TOC} = 0.5 \pm 1.7°C$. The statistical uncertainty of +/-1.7K is well below the thermal transition range of about 30K, usually determined by the anomalous step-like behavior of $\beta'_{thex,f}(T)$. $T_{g,TOC}$ includes the additional physical interpretation that at this temperature externally excited translational molecular motions in the sample are switched off.

For sure, at this point, the question arises to what extent the results obtained for *EP* can be generalized with reference to the thermo-optical coefficient and specifically to the TMOR method. Since the interaction of the sample with the refractometer prism and the temperature modulation play central roles in this $T_g$-determination technique, the utilization of the TMOR method is required. The currently available TMOR instruments are limited to a temperature range between -20°C and 128°C. Under these experimental boundary conditions, the glass transition temperature of the polymer sample to be investigated must lie in the temperature range between approx. 0°C and 70°C so that the measurement data for $\beta''_{add,f}(T)$ can still be approximated linearly with sufficient accuracy. This considerably narrows the current application range of the TMOR method and requires a significant extension of the temperature range of commercial TMOR instruments. Another limitation of the TMOR method results from the maximum temperature modulation frequency of about 30 mHz. The possibility to increase the maximum temperature modulation frequency by a factor of 10 would be desirable.

*In summary, by coupling the sample of interest to a refractometer prism, combined with a sinusoidal temperature modulation, one induces additional molecular modes of motion that are clearly reflected in the thermo-optical coefficient. This contribution to the thermo-optical coefficient vanishes at a well-defined temperature which can be defined as the glass transition temperature $T_{g,TOC}$.*

**Acknowledgements:**

The authors gratefully acknowledge the fruitful cooperation with Sandra Schlögl and David Reisinger from the Polymer Competence Center Leoben (PCCL) in Austria, as well as the discussions with Martine Philipp.

**Funding:**

The research was funded by the German Research Foundation (DFG), project number: 521902629.

**Data availability:**

The raw data required to reproduce these findings cannot be shared at this time as the data also forms part of an ongoing study. The processed data required to reproduce these findings are available to download from https://doi.org/10.5281/zenodo.10040655.